\documentstyle[12pt]{article}

\input{tcilatex}

\begin{document}

\title{Lepton MHD and magnetic field generation in SM \thanks{%
Talk given by author at the International Conference on Astrophysics and
High Energy Physics AHEP-03, Valencia, Spain, October 2003. Published in
Proceedings PRHEP-AHEP2003/059.} }
\author{V.B. Semikoz \\
IZMIRAN, Troitsk Moscow region 142190, Russia\\
e-mail address:semikoz@orc.ru }
\maketitle

\begin{abstract}
We derive a total set of MHD equations in SM describing evolution of a dense
plasma with neutrinos. First this is done for a hot pair plasma consisting
from electrons and positrons, neutrinos and antineutrinos of all flavors in
an isotropic medium like the early unverse plasma at the lepton stage. Then
we find how axial vector currents violating parity in SM contribute to MHD
for a slightly polarized (anisotropic) plasma where a new mechanism for the
amplification of mean magnetic fields arises due to the collective
neutrino-plasma interactions instead of assumed asymmetry of fluid velocity
vortices leading to the same effect of $\alpha^2$-dynamo.
\end{abstract}

\section{Introduction}

It is well-known that the magnetohydrodynamic (MHD) or macroscopic
description of a plasma is less detailed and much simpler than the kinetic
one which corresponds to the microscopic description of the plasma evolution
and therefore it is a more complicated approach.

The MHD equation system allows, in particular, to derive the Faradey
(induction) equation for magnetic fields in the standard model of
electroweak interactions (SM) including weak interaction terms. The main
goal of this work is the detailed derivation of Faradey equation in SM that
is important for the generation of primordial magnetic fields in cosmology
and magnetic fields in a supernova protostar where powerful neutrino fluxes
interact with the dense plasma.

We derive the full set of MHD equations using the standard method of moments 
\cite{Pitaevsky} for Relativistic Kinetic Equations (RKE) written in the
collisionless (Vlasov) approximation. There are other ways to derive MHD,
e.g. using the Lagrangian formalism for relativistic multicomponent fluid 
\cite{Brizard} while we prefer the method of the quantum RKE for lepton
plasma in SM \cite{Semikoz,OS} that is more approppriate to describe both
classical and spin properties of polarized plasmas permiated by an external
magnetic field.

Note that neutrino RKE is a useful tool to describe many phenomena in
astrophysics and cosmology. In particular, neutrinos play the most important
role for a supernova (SN) burst or in the lepton asymmetry formation before
the primordial nucleosynthesis in the early universe. The usual motivation
to use the RKE approach for neutrino propagation in a dense matter is
stipulated by the account of neutrino collisions: within a SN neutrinosphere
or in the hot lepton plasma of the early universe before neutrino decoupling.

However, in addition to collision integrals there are self-consistent weak
interaction terms in the neutrino RKE \cite{Semikoz} that are linear over
the Fermi constant $\sim G_F$ (see below section II) and analogous to the
Lorentz force terms for charge particles in the standard Boltzman RKE which
in turn are linear over the electric charge $\sim q$ ($q= - \mid e\mid$ for
electrons).

Let us remind that these self-consistent electromagnetic fields play a very
crucial role in the standard \ kinetics. In collisionless, or Vlasov
approximation, such kinetic equations describe, e.g. thermonuclear plasmas
in laboratory and stars for which an energy exchange between electromagnetic
waves propagating in plasma (=eigen modes) and charged particles
(wave-particle interaction) proceeds faster than via the direct particle
collisions (through particle-particle interaction) with all following issues
in collisionless plasma: instabilities, heating, etc.

One expects that the presence in the neutrino RKE\ of the similar terms $%
{\bf F}^{weak}({\bf x},t)\partial f^{(\nu )}({\bf k},{\bf x},t)/\partial 
{\bf k}$ where the weak force ${\bf F}^{weak}=\partial V^{weak}/\partial 
{\bf x}$ \ given by the neutrino interaction potential $V^{weak}\sim
G_{F}n_{e}({\bf x},t)$ is linear over $\sim G_{F}$ could lead for neutrinos
to some analogous collective interaction effects, e.g. to neutrino driven
streaming instability of plasma waves in an isotropic plasma \cite{RKE} or
instability of spin waves in a polarized medium \cite{OS}, and to the
generation of magnetic fields in hot plasma of early universe \cite%
{Brizard,SS}.

\bigskip

Note that in literature describing neutrino oscillation phenomena one
neglects the changes of neutrino momentum coming from non-forward
scattering, $\partial /\partial {\bf k(...)}$ =0, or refraction from density
variations, $\partial /\partial {\bf x(...)}$ =0, \cite{pantaleone}. Vice
versa, if one neglects neutrino masses and neutrino oscillations this weak
force remains as a main contribution in Vlasov approximation\ slightly
changing neutrino trajectories in the WKB$\ $approximation $kx\gg 1,$ where
the momentum transfer $\mid {\bf q\mid =}q\sim x^{-1}$obeys the WKB\
condition $q\ll k$. Such small momentum transfer  ${\bf q=k-k}^{\prime }\neq
0$ corresponds to the plasmon (\v{C}erenkov) emission by a massless ($%
K^{2}=k_{0}^{2}-k^{2}=K^{\prime 2}=k_{0}^{^{\prime }2}-k^{\prime 2}=0)$\
neutrino in medium, $\nu (K)\rightarrow \nu (K^{\prime })+\gamma ^{\ast }(Q),
$ and the resulting weak force ${\bf F}^{weak}$ is the friction force acting
on neutrino fluid due to this elementary process. Some details and the
explicite form of such force (acting from plasma on neutrinos) in dependence
on dispersion characteristics of an isotropic plasma can be found in the
preprint \cite{OS}.

In this work we try to find a new interesting consequence - generation of
magnetic fields by collective neutrino interactions following from the
presence of weak forces which are additive to the usual Lorentz force and
act from neutrino fluid on the electron-positron plasma. For that we should
derive the set of MHD equations and then to generalize Faradey equation in
SM.

\section{Lepton MHD in Standard Model (SM) of electroweak interactions}

In this Section we derive MHD equations using the method \cite{Pitaevsky} of
moments of kinetic equations , or integrating RKE's over momenta, $\int
d^3p(...)$, $\int d^3p{\bf {p}\times (...)}$, $\int d^3p\varepsilon_p\times
(...)$. We start from the simple case of unpolarized (isotropic) plasma and
in the next subsection we derive MHD using RKE's in a magnetized plasma \cite%
{OS}.

\subsection{Lepton MHD in unpolarized medium}

In an isotropic unpolarized plasma the collisionless RKE for electrons and
positrons ($e=\pm \mid e\mid $ with upper sign for positron) derived in SM
including weak forces takes the form \cite{RKE} 
\begin{eqnarray}  \label{RelKE}
&&\frac{\partial f^{(\pm )}({\bf p},{\bf x},t)}{\partial t}+{\bf v}\frac{%
\partial f^{(\pm )}({\bf p},{\bf x},t)}{\partial {\bf x}}\pm \mid e\mid
\left( {\bf E}({\bf x},t)+[{\bf v}\times {\bf B}({\bf x},t)]\right) \frac{%
\partial f^{(\pm )}({\bf p},{\bf x},t)}{\partial {\bf p}}\mp  \nonumber \\
&&\mp G_{F}\sqrt{2}\sum_{a}c_{V}^{(a)}\Bigl[-\nabla \left(n_{\nu _{a}}({\bf x%
},t)-n_{\bar{\nu}_{a}}({\bf x},t)\right)-\frac{\partial {\bf j}_{\nu_{a}} (%
{\bf x},t)-{\bf j }_{\bar{\nu}_{a}}({\bf x},t)}{\partial t}+  \nonumber \\
&&+{\bf v}\times \nabla \times ({\bf j}_{\nu_{a}}({\bf x},t)-{\bf j}_{\bar{%
\nu}_{a}}({\bf x },t))\Bigr]\frac{\partial f^{(\pm )}({\bf p},{\bf x},t)}{%
\partial {\bf p}}=0~,
\end{eqnarray}
where $j_{\mu }^{(\nu _{a},\bar{\nu}_{a})}({\bf x},t)=(n_{\nu_{a},\bar{\nu}%
_{a}}({\bf x},t),{\bf j}_{\nu_{a},\bar{\nu}_{a}}({\bf x},t))=\int
d^{3}k(k_{\mu }/E_{k})f^{(\nu ,\bar{\nu})}({\bf k},{\bf x},t)/(2\pi )^{3}$
is the neutrino (antineutrino) four-current density; $c_{V}^{(a)}=2\xi \pm
0.5$ is the vector coupling for $a=e,\mu ,\tau $ neutrinos with the upper
sign for the electron ones $a=e$.

We complete the system by the neutrino (antineutrino) collisionless RKE's: 
\begin{eqnarray}  \label{neutrino}
&&\frac{\partial f^{(\nu_a)}({\bf k}, {\bf x},t)}{\partial t} + {\bf n}\frac{
\partial f^{(\nu_a)}({\bf k}, {\bf x},t) }{\partial {\bf x}} + G_F\sqrt{2}%
c_V^{(a)}\left[- \nabla\left(n^{(e)}({\bf x},t) - n^{(\bar{e})}({\bf x}%
,t)\right)-\right.  \nonumber \\
&&\left.- \frac{\partial [{\bf j}^{(e)}({\bf x},t)- {\bf j}^{(\bar{e})}({\bf %
x},t)]}{\partial t} + {\bf n}\times \nabla\times \left({\bf j}^{(e)}({\bf x}%
,t) - {j}^{( \bar{e})}({\bf x},t)\right)\right]\frac{\partial f^{(\nu_a)}(%
{\bf k}, {\bf x},t)}{\partial {\bf k}} = 0~,  \nonumber \\
&& \frac{\partial f^{(\bar{\nu}_a)}({\bf {k}, {x},t)}}{\partial t} + {\bf n} 
\frac{\partial f^{(\bar{\nu}_a)}({k}, {x},t) }{\partial {\bf x}} - G_F\sqrt{2%
} c_V^{(a)}\left[- \nabla \left(n^{(e)}({\bf x},t) - n^{(\bar{e})}({\bf x}
,t)\right)-\right.  \nonumber \\
&&\left.- \frac{\partial [{\bf {j}^{(e)}({x},t)- {j}^{(\bar{e})}({x},t)]}}{
\partial t} + {\bf n}\times \nabla\times \left({\bf j}^{(e)}({\bf x},t) - 
{\bf j}^{( \bar{e})}({\bf x},t)\right)\right]\frac{\partial f^{(\bar{\nu}%
_a)}({\bf k}, {\bf x} ,t)}{\partial {\bf k}} = 0~,  \nonumber \\
&&
\end{eqnarray}
where $j_{\mu}^{(e,\bar{e})}({\bf x},t)=(n_{e,\bar{e}}({\bf x},t), {\bf j}%
_{e,\bar{e}}({\bf x},t))= \int d^3p(p_{\mu}/E_p)f^{(e,\bar{e})}({\bf p},{\bf %
x},t)/(2\pi)^3$ is the electron (positron) four-current density.

In 5-moment approximation of ideal hydrodynamics we neglect collisions and
hence omit viscosity, heat flux terms while retaining self-consistent
electroweak interactions between leptons. Thus we have to derive the
particle density conservation (continuity) equation, the motion (Euler)
equation (momentum conservation) and energy conservation equation. \vskip%
0.3cm {\it Continuity equations} \vskip0.3cm The weak interaction forces
above have the Lorentz structure or enter RKE's as well as the
electromagnetic Lorentz force in the third term of Eq. (\ref{RelKE}), $%
F_{j\mu }^{weak}({\bf x},t)\times (p^{\mu }/\varepsilon _{p})\partial
f^{(a)}({\bf p},{\bf x},t)/\partial p_{j}$. Hence they do not contribute to
the continuity equations \cite{RKE} which take the standard form $\partial
j_{\mu }^{(a)}/\partial x_{\mu }=0$ after integration of RKE's (\ref{RelKE})
and (\ref{neutrino}) over momenta $d^{3}p$, $d^{3}k$ correspondingly,
resulting in 
\begin{equation}
\frac{\partial n_{\pm }({\bf x},t)}{\partial t}+\frac{\partial \lbrack
n_{\pm }({\bf x},t)){\bf V}_{\pm }({x},t)]}{\partial {\bf x}}=0~,
\label{continuity}
\end{equation}%
for charged leptons and 
\begin{equation}
\frac{\partial n_{\nu _{a},\bar{\nu}_{a}}({\bf x},t)}{\partial t}+\frac{%
\partial \lbrack n_{\nu _{a},\bar{\nu}_{a}}({\bf x},t)){\bf V}^{(\nu _{a},%
\bar{\nu}_{a})}({\bf x},t))}{\partial {\bf x}}=0~,  \label{continuity1}
\end{equation}%
for neutrinos (antineutrinos). Here $n_{\pm }=n_{\pm }^{\prime }\gamma _{\pm
}$, $n_{\nu _{a},\bar{\nu}_{a}}=n_{\nu _{a},\bar{\nu}_{a}}^{\prime }\gamma
_{\nu _{a},\bar{\nu}_{a}}$ are the lepton densities in the laboratory
reference frame. The four-currents $j_{\mu }^{(a)}=n_{a}^{\prime }U_{\mu
}^{(a)}$ are given by the Lorentz-invariant densities $n_{a}^{\prime
}=j_{\mu }^{(a)}U^{(a)\mu }$ where $U^{(a)\mu }=(\gamma _{a},\gamma _{a}{\bf %
V}^{(a)})$ is the unit four-velocity of the plasma a-component, $U_{\mu
}^{(a)}U^{(a)\mu }=1$, $\gamma _{a}=(1-V_{a}^{2})^{-1/2}$, $a=\pm ,\nu _{a},%
\bar{\nu}_{a}$.

Considering the particular case of the hot pair plasma $T_-=T_+=T\gg \mu$
where the fast $e^{\pm}\gamma$ interaction provides equilibrium leading to
the zero chemical potentials $\mu_-=-\mu_+=\mu= 0$ and introducing the small
perturbations for comoving components ${\bf V}_{\pm}={\bf V} +\delta {\bf V}%
_{\pm}$, $\delta {\bf V}_{\pm}\ll{\bf V}$ that means $\gamma_-\approx
\gamma_+=\gamma $ we get the single continuity equation for the total
charged lepton density $n^{\prime}=n^{\prime}_- + n^{\prime}_+$ instead of
the two ones in Eq. (\ref{continuity}), 
\begin{equation}  \label{continuity2}
\frac{\partial \gamma n^{\prime}({\bf x},t)}{\partial t} + \frac{\partial
[\gamma n^{\prime}({\bf x},t)){\bf V}({\bf x},t)]}{\partial {\bf x}} =0~.
\end{equation}
Such hydrodynamical approximation means strong correlation in a dense plasma
between opposite charges due to which the continuous lepton medium becomes 
{\it electroneutral} conducting liquid (electrons and positrons move with
the same velocities as a whole) resulting in electric field vanishes while
magnetic field exists (lepton MHD, see below Eq. (\ref{Euler2})). Neglecting
protons the electroneutrality condition means that the background densities $%
n^{\prime}_{\pm 0}$ entering the total ones $n^{\prime}_{\pm}=n^{\prime}_{%
\pm 0}+ \delta n^{\prime}_{\pm}$ obey the equality 
\[
n^{\prime}_{-0}=n^{\prime}_{+0}=n_{0e}, 
\]
where in the hot plasma $n_{0e}=0.183 T^3$.

Note that the fluid (mean) velocity ${\bf V}$ differs from the microscopic $%
{\bf n}$ that enters the RKE of massless particles (\ref{neutrino}), $\mid 
{\bf n}\mid=1$. Of course, the Lorentz transformation with the unit vector $%
U_{\mu}^{(\nu_a)}=(\gamma_{\nu_a},\gamma_{\nu_a}{\bf V}_{\nu_a})$ does not
change the value of the microscopic four-momentum $k_{\mu}=(E_k, {\bf k})$, $%
E_k^2 - k^2=0$.

\vskip0.3cm {\it Motion equations} \vskip0.3cm Multiplying the RKE (\ref%
{RelKE}) by the momentum ${\bf p}$ and integrating it over $d^3p$ with the
use of the standard definitions of the fluid velocity\newline
${\bf V}_{\pm}({\bf x},t)= n^{-1}_{\pm}\int d^3p{v}f^{(\pm)}({\bf p}, {\bf x}
,t)/(2\pi)^3 $, the positron ( electron) density\newline
$n_{\pm}=\int d^3pf^{(\pm)}({\bf p}, {\bf x},t)/(2\pi)^3$, and the
generalized momentum of the lepton fluid ${\bf P}_{\pm}= w_{\pm}\gamma_{\pm}%
{\bf V}_{\pm}= n_{\pm}^{-1}\int d^3p{p}f^{(\pm)}({\bf p}, {\bf x}%
,t)/(2\pi)^3 $, one obtains the Euler equation that coincides with Eq. (4.6)
derived in \cite{Brizard} using another (relativistic Lagrangian) approach
for multicomponent fluid, 
\begin{eqnarray}  \label{Euler}
&&\left(\partial_t + {\bf V}_{\pm}\cdot \nabla \right){\bf P}_{\pm} = - 
\frac{\nabla p_{\pm }^{\prime}}{n_{\pm}} \pm \mid e\mid\left({\bf E} + [{\bf %
V}_{\pm}\times {\bf B}]\right) \mp  \nonumber \\
&& \mp G_F\sqrt{2}\sum_{\nu_a} c_V^{a}\left[- \nabla \delta n_{\nu_a}({\bf x 
},t) - \frac{\partial \delta {\bf j}_{\nu_a}({\bf x},t)}{\partial t} + {\bf V%
}_{\pm}\times \nabla\times \delta {\bf j}_{\nu_a}({\bf x},t) \right]~,
\end{eqnarray}
where $\delta n_{\nu_a}= n_{\nu_a} - n_{\bar{\nu}_a}$, $\delta {\bf j}%
_{\nu_a}= {\bf j}_{\nu_a} - {\bf j}_{\bar{\nu}_a}$ are the neutrino density
and neutrino 3-current density asymmetries respectively; $w_{\pm}=e_{\pm} +
p_{\pm}^{\prime}/n_{\pm}^{\prime}$ is the Lorentz-scalar enthalpy per one
particle; $e_{\pm}$, $p_{\pm}^{\prime}=n_{\pm }^{\prime}T_{\pm}$ are the
internal energy and the pressure correspondingly, $T_{\pm}$ is the
Lorentz-invariant temperature. In particular, for the J\"{u}ttner
equilibrium distribution $f_{\pm}^{eq}(p) =\exp [(\mu_{\pm} -
p_{\mu}U^{\mu})/T_{\pm}]$, where $\mu_{\pm}$ is the Lorentz-invariant
chemical potential, the thermodynamical characteristics are also
Lorentz-invariant, 
\begin{equation}  \label{thermo}
w_{\pm}=m_e\frac{K_3(m_e/T_{\pm})}{K_2(m_e/T_{\pm})},~~~ e_{\pm}=w_{\pm} -
T_{\pm},~~~ p_{\pm}^{\prime}=4\pi m_e^2T_{\pm}^2K_2(m_e/T_{\pm})\exp
(\mu_{\pm}/T_{\pm})~.
\end{equation}
For equilibrium pair plasma $T_+=T_-=T$ all these characteristics coincide, $%
w_+=w_-=w_e$, $p_+^{\prime}=p_-^{\prime}=p_e$, etc.

Summing Euler equations for electrons and positrons (\ref{Euler}) one
obtains the motion equation (electric field and neutrino density terms do
not contribute) 
\begin{eqnarray}  \label{e+e-}
&&\frac{{\rm d}({\bf P}_+ + {\bf P}_-)}{{\rm d}t} = - \frac{\nabla
(p_+^{\prime}+ p_-^{\prime})}{\gamma n_{0e}} + \mid e\mid({\bf V}_+ - {\bf V}%
_-)\times {\bf B} -  \nonumber \\
&& - G_F\sqrt{2}\sum_{\nu_a}c_V^{(a)}\left[({\bf V}_+ - {\bf V}_-)\times
\nabla \times \delta {\bf j}_{\nu_a}({\bf x},t)\right]~,
\end{eqnarray}
where ${\rm d}/{\rm d}t= \partial/\partial t + {\bf V}\cdot \nabla$.

Then we use in (\ref{e+e-}) the Maxwell equation without displacement
current ($\partial {\bf E}/\partial t$ is omitted in MHD ), $\delta {\bf j}%
_e^{(em)}=\mid e\mid n_{0e}\gamma ({\bf V}_+ -{\bf V}_-)= 2\mid e\mid
n_e\delta {\bf V}= {\rm rot}~{\bf B}/4\pi$ where $n_e=\gamma n_{0e}$ is the
plasma density in the laboratory reference frame. We put also the total
pressure $p=p_+^{\prime}+ p_-^{\prime}=2p_e$, the total enthalpy $w=w_+ +
w_-=2w_e$ introducing the total generalized momentum ${\bf P}={\bf P}_+ +%
{\bf P}_-=w\gamma {\bf V}$.

Thus, we obtain finally the MHD motion equation for pairs generalized in SM
with neutrinos, 
\begin{equation}  \label{Euler2}
\frac{{\rm d}{\bf P}}{{\rm d}t} = - \frac{\nabla p}{n_{e}} + \frac{ {\rm rot}%
{\bf B}\times {\bf B}}{4\pi n_e}- \frac{G_F\sqrt{2}}{\mid e\mid 4\pi n_e}%
\sum_{\nu_a}c_V^{(a)} \left[{\rm rot}{\bf B}\times \nabla \times \delta {\bf %
j}_{\nu_a}({\bf x},t)\right]~,
\end{equation}

The motion equations for neutrinos and antineutrinos are derived multiplying
the RKE (\ref{neutrino}) by the momentum ${\bf {k}}$ and integrating over $%
d^3k$, 
\begin{eqnarray}  \label{nuEuler}
&&\frac{{\rm d}{\bf K}_{\nu_a}}{{\rm d}t} = - \frac{\nabla p_{\nu_a}^{\prime}%
}{n_{\nu_a}} +{\bf F}_{\nu_a},  \nonumber \\
&& \frac{{\rm d}{\bf K}_{\bar{\nu}_a}}{{\rm d}t} = - \frac{\nabla p_{ \bar{%
\nu}_a}^{\prime}}{n_{\bar{\nu}_a}} +{\bf F}_{\bar{\nu}_a},
\end{eqnarray}
where the generalized momenta 
\[
{\bf K}_{\nu_a,\bar{\nu}_a}= \gamma_{\nu_a,\bar{\nu}_a}w_{\nu_a,\bar{\nu}_a} 
{\bf V}_{\nu_a,\bar{\nu}_a}= n_{\nu_a,\bar{\nu}_a}^{-1}\int d^3k{k}f^{(\nu_a,%
\bar{\nu}_a)} ({\bf k}, {\bf x},t)/(2\pi)^3~, 
\]
are given by the Lorentz-invariant thermodynamical functions $w_{\nu_a,\bar{
\nu}_a}= e_{\nu_a,\bar{\nu}_a} + p_{\nu_a,\bar{\nu}_a}^{\prime}/n_{\nu_a, 
\bar{\nu}_a}^{\prime}$; the weak forces ${\bf F}_{\nu}$ given by 
\begin{eqnarray}  \label{force}
&&{\bf F}_{\nu_a}= + G_F\sqrt{2}c_V^{(a)}\Bigl[ - \nabla \delta n^{(e)}({\bf %
x},t) -\frac{\partial \delta {\bf j}^{(e)}({\bf x},t)}{\partial t} + {\bf V}%
_{\nu_a}\times \nabla\times \delta {\bf j}^{(e)}({\bf x},t)\Bigr],  \nonumber
\\
&& {\bf F}_{\bar{\nu}_a}= - G_F\sqrt{2}c_V^{(a)} \Bigl[- \nabla\delta
n^{(e)}({\bf x},t) -\frac{\partial \delta {\bf j}^{(e)}({\bf x},t)}{\partial
t} + {\bf V}_{ \bar{\nu}_a}\times \nabla\times \delta {\bf j}^{(e)}({\bf x}%
,t)\Bigr]~,  \nonumber \\
\end{eqnarray}
have opposite signs and, in general, depend on {\it different} fluid
velocities, ${\bf V}_{\nu_a}\neq {\bf V}_{\bar{\nu}_a}$. Here we input
charged lepton density and 3-current density asymmetries, $\delta n^{(e)} =
n_- - n_+ $, $\delta {\bf j}^{(e)}={\bf j}^{(e)}- {\bf j}^{(\bar{e})} $
which are small in hot plasma.

Since there are different fluid velocities as well as possible different
thermodynamical functions, $w_{\nu_a}=e_{\nu_a} + T_{\nu_a}=4T_{\nu_a}\neq
w_{\bar{\nu}_a}= 4T_{\bar{\nu}_a}$ with the equation of state for massless
neutrinos, $p_{\nu}=e_{\nu}/3$ (see (\ref{thermo}) for massless particles $%
m_{\nu}=0$), we consider different motion equations for neutrinos and
antineutrinos (\ref{nuEuler}). Note that the inequality for electron
neutrino species, $T_{\nu_e}\neq T_{\bar{\nu}_e}$ can arise due to
beta-processes and the CC-current interaction and leads to a temperature
difference for electron and muon (tau) neutrino components. For the latter ($%
\nu_{\mu}$, $\bar{\nu}_{\mu}$ ) one expects same temperatures, however, we
do not use this property to simplify the system (\ref{nuEuler}). \vskip0.3cm 
{\it Energy equations} \vskip0.3cm Multiplying the RKE (\ref{RelKE}) by the
energy $E_p$ and integrating over $d^3p$ one obtains the energy conservation
law (upper sign for positrons) 
\begin{eqnarray}  \label{energy}
&&\frac{\partial [\gamma_{\pm}^2n_{\pm}^{\prime}e_{\pm} +\gamma_{\pm}^2 {\bf %
V}^2_{\pm}p_{\pm}^{\prime}]}{\partial t} + \frac{\partial
[\gamma_{\pm}^2n_{\pm}^{\prime}w_{\pm}{\bf V}_{\pm}]}{\partial {\bf x}} = 
\nonumber \\
&&=\pm \gamma_{\pm}n_{\pm}^{\prime}{\bf V}_{\pm} \cdot \left(\mid e\mid {\bf %
E} - {\bf F}^{weak}_e\right)~,
\end{eqnarray}
where the inner energy $e_{\pm}$, the enthalpy $w_{\pm}$, the pressure $%
p_{\pm}$ are given by Eq. (\ref{thermo}); the weak force in the r.h.s.
acting on charged leptons is given by 
\begin{equation}  \label{weakforce}
{\bf F}^{weak}_e= G_F\sqrt{2}\sum_{\nu_a} c_V^{a}\left[- \nabla \delta
n_{\nu_a}({\bf x},t) - \frac{\partial \delta {\bf j}_{\nu_a}({\bf x},t)} {%
\partial t} \right]~.
\end{equation}

Adding energy equations (\ref{energy}) and using the relation ${\bf E}=-{\bf %
V} \times {\bf B}$ that is valid for an ideal conducting medium one gets the
MHD energy equation for pairs generalized here in SM including weak forces, 
\begin{equation}  \label{energy1}
\frac{\partial [\gamma^2n^{\prime}e_e +\gamma^2 {\bf V}^2p]}{\partial t} + 
\frac{\partial [\gamma^2n^{\prime}w_e{\bf V}]}{\partial {\bf x}} = - \frac{(%
{\rm rot}~{\bf B}) \cdot \Bigl(\mid e\mid[{\bf V}\times {\bf B}] + {\bf F}%
^{weak}_e\Bigr) }{4\pi \mid e\mid }~,
\end{equation}
where the force ${\bf F}^{weak}_e$ is given by (\ref{weakforce}), $%
n^{\prime}= n_+^{\prime}+ n_-^{\prime}$, ${\bf V}_-={\bf V}_+ \approx {\bf V}
$, $\gamma_+=\gamma_-\approx \gamma$ and meaning the equilibrium reached
through the fast $e\gamma$-interaction, $T_+=T_-=T$ we put $e_e= e_+ = e_-$, 
$w_e= w_+ = w_-$, $p= p_+^{\prime}+ p_-^{\prime}$ in the agreement with (\ref%
{thermo}).

Analogously multiplying the neutrino (antineutrino) RKE (\ref{neutrino}) by
the energy $E_k$ and integrating over $d^3k$ one obtains the energy equation
(upper sign for neutrinos) 
\begin{eqnarray}  \label{energy2}
&&\frac{\partial [\gamma_{\nu_a,\bar{\nu}_a}^2 n_{\nu_a,\bar{\nu}
_a}^{\prime}e_{\nu_a,\bar{\nu}_a} +\gamma_{\nu_a,\bar{\nu}_a}^2 {\bf V}%
^2_{\nu_a,\bar{\nu}_a}p_{\nu_a,\bar{\nu}_a}^{\prime}]}{\partial t} + \frac{
\partial [\gamma_{\nu_a,\bar{\nu}_a}^2n_{\nu_a,\bar{\nu}_a}^{\prime}w_{%
\nu_a, \bar{\nu}_a}{\bf V}_{\nu_a,\bar{\nu}_a}]}{\partial {\bf x}} = 
\nonumber \\
&&= \pm\gamma_{\nu_a,\bar{\nu}_a} n_{\nu_a,\bar{\nu}_a}^{\prime} ({\bf V}%
_{\nu_a,\bar{\nu}_a} \cdot {\bf F}_{\nu_a,\bar{\nu}_a})~,
\end{eqnarray}
where the weak force acting on neutrinos from the pair plasma ${\bf F}%
_{\nu_a,\bar{\nu}_a}$ is given by Eq. (\ref{force}).

The set of MHD equations: the continuity ones (\ref{continuity1}), (\ref%
{continuity2}), the motion ones (\ref{Euler2}), (\ref{nuEuler}) and the
energy ones, (\ref{energy1}), (\ref{energy2}) is completed by the Faradey
equation for the magnetic field ${\bf B}$ generalized in SM due to weak
interactions (see next section, Eq. (\ref{Faradey})).

\subsection{Lepton MHD in polarized medium}

Analogously with the case of unpolarized medium we can derive MHD equations
in the presence of a strong {\it large-scale} uniform magnetic field ${\bf B}%
_0$ which polarizes plasma populating partially the main Landau
(non-degenerate) levels for free electrons and positrons. Other levels being
populated by leptons with opposite spin projections are degenerate (the
factor Lande $g_e=2$ doubles such states) and do not contribute to the
medium polarization.

The lepton density at the main Landau level in {\it anisotropic medium} is
given by 
\begin{equation}  \label{polar}
n_0^{(\pm)}= \int \frac{d^3p}{(2\pi)^3}S_0^{(\pm)}(\varepsilon_p) =\frac{
\mid e\mid B_0}{2\pi^2}\int_0^{\infty}dpf_0^{(\pm)}(\varepsilon_p)~,
\end{equation}
where in the hot plasma $T\gg \mu$ one obtains $n_0^{(\pm)}\simeq \mid e\mid
B_0T\ln 2/2\pi^2$.

Now using the electron RKE Eq. (30) from \cite{OS} we can generalize the
Euler equation for electrons and positrons (\ref{Euler}) for the case of a
polarized medium, 
\begin{eqnarray}  \label{Eulerpol}
&&\left(\partial_t + {\bf V}_{\pm}\cdot \nabla \right){\bf P}_{\pm} = - 
\frac{\nabla p_{\pm }^{\prime}}{n_{\pm}} \pm \mid e\mid\left({\bf E} + [{\bf %
V}_{\pm}\times {\bf B}]\right) \mp  \nonumber \\
&& \mp G_F\sqrt{2}\sum_{\nu_a} c_V^{a}\left[ - \nabla \delta n_{\nu_a}({\bf x%
},t) - \frac{\partial \delta {\bf j}_{\nu_a}({\bf x},t)} {\partial t} + {\bf %
V}_{\pm}\times \nabla \times \delta {\bf j}_{\nu_a}({\bf x},t) \right] \mp 
\nonumber \\
&& \mp \frac{G_F\sqrt{2}}{n_e}\sum_{\nu_a}c_A^{(a)}\left[ n_0^{(\pm)}\hat{%
{\bf b}}^{(0)}{\rm div}~\delta {\bf j}_{\nu_a}({\bf x},t) -
N_0^{(\pm)}\nabla (\hat{{\bf b}}^{(0)} \cdot \delta {\bf j}_{\nu_a}({\bf x}%
,t)\right] ~,
\end{eqnarray}
where in a non-relativistic (NR) plasma, the relativistic polarization
density terms 
\[
N_0^{(\pm)}= \frac{n_0^{\pm}}{3} + \frac{\mid e\mid B_0m_e}{18\pi^2}
\int_0^{\infty}f_0^{\pm}(\varepsilon_p)dp \frac{\partial v(3 -v^2)}{\partial
p} 
\]
coincide with the main Landau level contributions, $N_0^{(\pm)}\to
n_0^{(\pm)}$, given by Eq. (\ref{polar}); ${\bf B}= {\bf B}_0 + {\bf B}%
^{\prime}$ is the total magnetic field.

Adding equations (\ref{Eulerpol}) we obtain finally the pair motion equation
in polarized medium: 
\begin{eqnarray}  \label{Euler3}
&&\frac{{\rm d}{\bf P}}{{\rm d}t} = - \frac{\nabla p}{n_{e}} + \frac{ {\rm %
rot}{\bf B}\times {\bf B}}{4\pi n_e}- \frac{G_F\sqrt{2}}{\mid e\mid 4\pi n_e}%
\sum_{\nu_a}c_V^{(a)} \left[{\rm rot}{\bf B}\times \nabla \times \delta {\bf %
j}_{\nu_a}({\bf x},t)\right] +  \nonumber \\
&& + \frac{G_F\sqrt{2}}{n_e}\sum_ac_A^{(a)}\Bigl[(n_0^{(-)} - n_0^{(+)})\hat{
{\bf b}}^{(0)}{\rm div}~\delta {\bf j}_{\nu_a}({\bf x},t) -  \nonumber \\
&&- (N_0^{(-)} - N_0^{(+)})\nabla (\hat{{\bf b}}^{(0)} \cdot \delta {\bf j}%
_{\nu_a}({\bf x},t))\Bigr]~.  \nonumber \\
\end{eqnarray}

Note that the polarization asymmetries $n^{(-)}_0 - n^{(+)}_0$, $N_0^{(-)}-
N_0^{(+)}$ are small in the hot relativistic plasma of early universe while
in a degenerate electron gas of a magnetized supernova, $T\ll \mu$, these
asymmetries can be large since $n_0^{(-)}=\mid e\mid B_0\mu/2\pi^2\gg
n_0^{(+)}=(\mid e\mid B_0T/2\pi^2)e^{- \mu/T}$.

Assuming ${\bf {B}^{\prime}\ll {B}_0}$ we can include the perturbative field 
${\bf {B}^{\prime}({x},t)}$ into the polarization terms in the last lines of
Eq. (\ref{Euler3}) with the change ${\bf B}_0\to {\bf B}$, $\hat{{\bf b}}%
^{(0)}\to \hat{{\bf b}}$.

In general, one can consider the limit of strong magnetic fields (or diluted
media) for which the main Landau level is populated only. E.g. a degenerate
electron gas obeying the condition $eB\geq \mu^2/2$ would be fully
polarized, or $n_e\approx n_0^{(-)}$ \cite{Hiroshi}, that could lead to
comparable contributions of pseudovector and vector terms in the pair motion
equation (\ref{Euler3}).

The neutrino (antineutrino) motion equations take the form which is similar
to Eq. (\ref{nuEuler}) while in a polarized medium the vector force for
neutrinos ${\bf F}_{\nu_a}$ (\ref{force}) (and similarly ${\bf F}_{\bar{ \nu}%
_a}$ for antineutrinos) is added with the additional axial vector force, $%
{\bf F}_{\nu_a}\to {\bf F}_{\nu_a} + {\bf F}_{\nu_a}^{(A)}$, 
\begin{eqnarray}  \label{nuEuler2}
&&\frac{{\rm d}{\bf K}_{\nu_a}}{{\rm d}t} = - \frac{\nabla p_{\nu_a}^{\prime}%
}{n_{\nu_a}} +{\bf F}_{\nu_a} + {\bf F}_{\nu_a}^{(A)},  \nonumber \\
&& \frac{{\rm d}{\bf K}_{\bar{\nu}_a}}{{\rm d}t} = - \frac{\nabla p_{ \bar{%
\nu}_a}^{\prime}}{n_{\bar{\nu}_a}} +{\bf F}_{\bar{\nu}_a} + {\bf F}_{\bar{\nu%
}_a}^{(A)}.
\end{eqnarray}
The latter term ( ${\bf F}_{\nu_a}^{(A)}$ and similarly ${\bf F}_{\bar{\nu}%
_a}^{(A)}$ with the change of common sign and fluid velocity ${\bf V}%
_{\nu_a}\to {\bf V}_{\bar{\nu}_a} $), 
\begin{equation}  \label{spin}
{\bf F}_{\nu_a}^{(A)}=\frac{G_Fc_A^{(a)}}{\sqrt{2}}\left[ - \nabla \delta
A_0({\bf x},t) - \frac{\partial \delta {\bf A}({\bf x},t )}{\partial t} + 
{\bf V}_{\nu_a}\times \nabla \times \delta {\bf A}({\bf x},t)\right]~,
\end{equation}
depends on the spin density asymmetry $\delta A_{\mu}({\bf x},t)$, 
\begin{eqnarray}
&&\delta A_{\mu}({\bf x},t)= A_{\mu}^{(-)}({\bf x},t) - A_{\mu}^{(+)}({\bf x}%
,t)~,  \nonumber \\
&&A_{\mu}^{(\pm)}({\bf x},t)= m_e\int \frac{d^3p}{(2\pi)^3} \frac{1}{%
\varepsilon_p} \left(\frac{{\bf p}{\bf S}^{(\pm)}({\bf p}, {\bf x}, t)}{m_e}%
; {\bf S}^{(\pm)}({\bf p} , {\bf x}, t) + \frac{{\bf p}({\bf p} \cdot{\bf S}%
^{(\pm)}({\bf p}, {\bf x}, t))}{m_e(\varepsilon_p + m_e)}\right)~,  \nonumber
\\
\end{eqnarray}
that is given by the charged lepton spin distributions ${\bf S}^{(\pm)}({\bf %
p}, {\bf x}, t)$ obeying the spin RKE like Eq. (12) in \cite{OS}. In NR
plasma such electron spin distribution defines the well-known hydrodynamical
characteristic - magnetization ${\bf m}({\bf x},t)=\mid \mu_B\mid \int
(d^3p/(2\pi)^3 {\bf S}^{(-)}({\bf p}, {\bf x}, t) \approx \mid \mu_B\mid 
{\bf A}({\bf x},t)$ which obeys the Bloch evolution equation and completes
the system of MHD equations for fermions in a polarized medium \cite{OS}.

We note here that the neutrino (antineutrino) currents ${\bf j}_{\nu_a,\bar{
\nu}_a}({\bf x},t)$ entering through weak forces the pair motion equation (%
\ref{Euler3}) are connected with the neutrino generalized momenta ${\bf K}%
_{\nu_a,\bar{\nu}_a}({\bf x},t)$ via 
\begin{equation}  \label{relation}
{\bf j}_{\nu_a,\bar{\nu}_a}({\bf x},t)= \frac{n^{\prime}_{\nu_a,\bar{\nu}_a}%
}{w_{\nu_a,\bar{\nu}_a}} {\bf K}_{\nu_a,\bar{\nu}_a}({\bf x},t).
\end{equation}
Note also that continuity equations (\ref{continuity}), (\ref{continuity1})
are fulfilled in polarized medium \cite{OS}. We do not consider here energy
equations that are easily derived from RKE's in polarized medium analogously
to Eqs. (\ref{energy}-\ref{energy2}).

\section{Faradey equation in SM}

In order to derive Faradey equation let us multiply the electron (positron)
hydrodynamical motion equation (\ref{Eulerpol}) by $-\mid e\mid$ (and +$\mid
e\mid$) correspondingly and then sum them to obtain the auxiliary result for
the electric field in a polarized plasma: 
\begin{eqnarray}  \label{electric}
&&{\bf E}= - \frac{1}{2}\sum_{\sigma=\pm}{\bf V}_{\sigma}\times {\bf B} +
\sum_{\sigma=\pm }\frac{e_{\sigma}}{2e^2}\left(\partial_t{\bf P}_{\sigma} +
\nu_{em}\delta {\bf P}_{\sigma } + ({\bf V}_{\sigma}\nabla){\bf P}%
_{\sigma}\mp (V_{\sigma})_n \nabla (P_{\sigma})_n\right) +  \nonumber \\
&& + \frac{G_F\sqrt{2}}{\mid e\mid}\sum_{\nu_a}c_V^{(a)} \Bigl[ - \nabla
\delta n_{\nu_a} - \partial_t \delta {\bf j}_{\nu_a} + \frac{1}{2}
\sum_{\sigma=\pm}{\bf V}_{\sigma}\times \nabla\times \delta {\bf j}_{\nu_a} %
\Bigr]-  \nonumber \\
&&- \frac{G_F}{\sqrt{2}\mid e\mid n_e}\sum_{\nu_a}c_A^{(a)}\left[(n_0^{(-)}
+ n_0^{(+)}) \hat{{\bf b}}~\frac{\partial \delta n_{\nu_a}( {\bf x},t)}{%
\partial t} +\right.  \nonumber \\
&&\left. + (N_0^{(-)} + N_0^{(+)})\nabla (\hat{{\bf b}} \cdot \delta {\bf j}%
_{\nu_a}({\bf x},t))\right]~.
\end{eqnarray}

Let us stress that instead of the {\it difference} of electron and positron
contributions in axial vector terms entering the pair motion equation (\ref%
{Euler3}) and given by the polarized density asymmetries $\sim (n_0^{(-)}-
n_0^{(+)})$ we obtained here the {\it sum} of them $\sim (n_0^{(-)}+
n_0^{(+)})$ that can lead to an essential effect in hot plasma (see below
section IV).

Using for the last term at the first line of Eq. (\ref{electric}) the
identity $(V_{\sigma})_n\nabla (P_{\sigma})_n - ({\bf V}_{\sigma}\nabla){\bf %
P}_{\sigma} = {\bf V}_{\sigma}\times \nabla \times {\bf P}_{\sigma}$ and the
thermodynamics relation for the work $dR_{\sigma}/dt ={\bf V}_{\sigma}{\rm d}%
{\bf P}_{\sigma}/{\rm d}t= - p_{\sigma}{\rm d}v_{\sigma}/{\rm d}t$, $%
(V_{\sigma})_n\nabla (P_{\sigma})_n = \nabla (\varepsilon_{\sigma} -
T_{\sigma}S_{\sigma}) + S_{\sigma}\nabla T_{\sigma}$, where $%
\varepsilon_{\sigma}~,S_{\sigma}$ are the internal energy and entropy per
one particle (of the kind $\sigma=\pm$), $p_{\sigma},~v_{\sigma}$, $%
T_{\sigma}$ are the pressure, the volume and the temperature
correspondingly; then substituting Eq. (\ref{electric}) into the Maxwell
equation $\partial_t {\bf B} = - \nabla \times {\bf E}$ we obtain the
Faradey equation generalized in SM with neutrinos and antineutrinos: 
\begin{eqnarray}  \label{Faradey}
&&\partial_t {\bf B} =\nabla\times {\bf V}\times {\bf B} - \nabla\times
\eta\nabla\times {\bf B} + \sum_{\sigma}\left(\frac{e_{\sigma}}{2e^2}\right)
\nabla T_{\sigma}\times \nabla S_{\sigma}-  \nonumber \\
&& -\sum_{\sigma}\left(\frac{e_{\sigma}}{2e^2}\right)\nabla \times
\left(\partial_t {\bf P}_{\sigma} - {\bf V}_{\sigma}\times \nabla \times 
{\bf P}_{\sigma}\right)-  \nonumber \\
&&- \frac{G_F\sqrt{2}}{\mid e\mid}\sum_{\nu_a} c_V^{(a)}\nabla \times
\left(\partial_t \delta {\bf j}_{\nu_a} - {\bf V}\times \nabla \times \delta 
{\bf j}_{\nu_a}\right) +  \nonumber \\
&& + \frac{G_F\sqrt{2}}{2\mid e\mid}\sum_{\nu_a}c_A^{(a)}\left[\nabla\times
\left(\frac{n_0^{(-)} + n_0^{(+)}}{n_e}\right)\left(\hat{{\bf b}}\frac{%
\partial \delta n_{\nu_a}}{\partial t}\right) + \right.  \nonumber \\
&&\left. + \nabla\times \left(\frac{N_0^{(-)} + N_0^{(+)}}{n_e}%
\right)\nabla( \hat{{\bf b}} \cdot \delta {\bf j}_{\nu_a}) \right] ~. 
\nonumber \\
\end{eqnarray}
Here the equalities $\delta {\bf V}_+ +\delta {\bf V}_-=0$, or ${\bf V}_+ +%
{\bf V}_-=2{\bf V}$, ${\bf V}_+ -{\bf V}_-= 2\delta {\bf V}_+\equiv 2\delta 
{\bf V}$ followed from the $e\gamma$-equilibrium are taken into account; the
magnetic diffusion coefficient $\eta=(4\pi \sigma_{cond})^{-1}$ stems from
the third term in the electric field (\ref{electric}) given by the
electromagnetic collision frequency $\nu_{em}$, which enters the plasma
conductivity $\sigma_{cond}=\omega_p^2(\varepsilon_e/w_e)/4\pi \nu_{em}$
with $\varepsilon_e$, $\omega_p=\sqrt{4\pi\alpha n_e/\varepsilon_e}$ being
the internal energy and the plasma frequency correspondingly. In the
non-relativistic plasma the enthalpy $w_e$ coincides with the internal
energy, $w_e\approx \varepsilon_e\approx m_e$, while in the hot relativistic
plasma $w_e=4T$, $\varepsilon_e=3T$. For the uniform conductivity the second
term takes the standard form $+ \eta\nabla^2{\bf B}$.

The first term in the r.h.s. (\ref{Faradey}) represents the nonlinear dynamo
effect, the third one is the Biermann battery effect. The fourth term can be
neglected for small fluctuations $\delta {\bf P}\ll {\bf P}$, $\delta {\bf V}%
\ll {\bf V}$.

In an unpolarized medium we can omit all terms in last lines which are
proportional to the axial vector coupling $c_A^{(a)}$. The remaining
standard terms and weak interaction vector terms ($\sim c_V^{(a)}$)
reproduce the Faradey equation (5.7) in \cite{Brizard} for the {\it ideal}
pair plasma ($\eta=0$) interacting with neutrinos (antineutrinos).

The neutrino (antineutrino) currents ${\bf j}_{\nu_a,\bar{\nu}_a}$ entering (%
\ref{Faradey}) are given in Eq. (\ref{relation}) by their generalized
momenta ${\bf K}_{\nu_a,\bar{\nu}_a}$ which in turn obey the motion
equations (\ref{nuEuler}), (\ref{nuEuler2}).

In the next section we consider an application \footnote{%
Results in section below were obtained together with D.D. Sokoloff in \cite%
{SS}.} of the generalized Faradey equation (\ref{Faradey}) which includes
weak interaction terms violating parity to the actual problem of magnetic
field generation in the early universe plasma.

\section{Large-scale magnetic field generation in early universe}

The main problem of primordial magnetic field generation that leads to a
seed of observable galactic magnetic fields is an inconsistency of their
values $B$ and correlation lengths $L_0$ obtained in the different scenarios.

There are many ways how to generate small-scale random magnetic fields with
large values of $B_{rms}=\sqrt{<B^2>}$, e.g. using some causal mechanisms
like bubble collisions at phase transitions, while the correlation length of
such magnetic fields evolved (via inverse cascade) during expansion of
universe into large-scale magnetic fields turns out to be too small at
present time, $L_0\sim tens~parsecs$, to reach the size $L_0\sim 100~kps$
for galactic magnetic field, or even more ($\gg$ Mps) for extragalactic
magnetic fields. The other way using inflation scenario allows, vice versa,
to get large-scale (a few Mps) magnetic fields while their strength occurs
too small for observable magnetic fields.

Let us simplify the Faradey equation (\ref{Faradey}) rewriting it as a
simple governing equation for mean magnetic field evolution 
\begin{equation}
\frac{\partial {\bf B}}{\partial t}=\nabla \times \alpha {\bf B}+\eta \nabla
^{2}{\bf B}~,  \label{Faradey2}
\end{equation}%
where we omitted: dynamo term neglecting any macroscopic rotation in plasma
of early universe, Biermann battery effect and weak interaction terms given
by the vector coupling $c_{V}^{(a)}$ because of the absence of neutrino
vorticity in isotropic neutrino gas, $\nabla \times {\bf j}_{\nu
_{a}}(r,t)=0.$ The neutrino fluid vorticity vanishes also for isotropic
neutrino emission from a supernova since in the diffusion approximation
neutrino fluxes (not a particular neutrino) propagate along radii, ${\bf j}%
_{\nu _{a}}(r,t)\parallel {\bf r}$ even under neutrinosphere${\bf .}$

In \ Eq. (\ref{Faradey2}) we approximate the tensor $\alpha _{ij}$ coming in 
${\bf E}$ from the axial vector force in (\ref{Eulerpol}) by the first
diagonal ($\sim \alpha \delta _{ij}$) term: 
\begin{eqnarray}
&&\alpha =\frac{G_{F}}{2\sqrt{2}\mid e\mid B}\sum_{a}c_{e\nu _{a}}^{(A)}%
\left[ \left( \frac{n_{0}^{(-)}+n_{0}^{(+)}}{n_{e}}\right) \frac{\partial
\delta n_{\nu _{a}}}{\partial t}\right] \simeq   \nonumber  \label{helicity}
\\
&\simeq &\frac{\ln 2}{4\sqrt{2}\pi ^{2}}\left( \frac{10^{-5}T}{%
m_{p}^{2}\lambda _{{\rm fluid}}^{(\nu )}}\right) \left( \frac{\delta n_{\nu }%
}{n_{\nu }}\right) ~,
\end{eqnarray}%
where densities $n_{0}^{(\pm )}$ are given by Eq. (\ref{polar}), $n_{\nu
}/n_{e}=0.5$, and we assume a scale of neutrino fluid inhomogeneity $t\sim
\lambda _{{\rm {fluid}}}^{(\nu )}$, that is small comparing with a large $%
\Lambda $-scale of the mean magnetic field ${\bf B}$, $\lambda _{{\rm fluid}%
}^{(\nu )}\ll \Lambda $.

The diffusion coefficient $\eta \approx (4\pi 137~T)^{-1}$ is given by the
relativistic plasma conductivity.

For a small neutrino chemical potential $\mu_{\nu}$, $\xi_{\nu_a}(T)=\mu_{%
\nu_a}(T)/T\ll 1$, the neutrino asymmetry in the r.h.s. of Eq. (\ref%
{helicity}) is the algebraic sum following the sign of the axial coupling, $%
c^{(A)}_{e\nu_a}= \pm 0.5$, 
\begin{equation}  \label{asymmetry}
\frac{\delta n_{\nu}}{n_{\nu}}\equiv \sum_ac^{(A)}_{e\nu_a}\frac{\delta
n_{\nu_a}}{n_{\nu_a}}= \frac{2\pi^2}{9\zeta(3)}[\xi_{\nu_{\mu}}(T) +
\xi_{\nu_{\tau}}(T) - \xi_{\nu_e}(T)]~.
\end{equation}

We stress that the Eq.~(\ref{Faradey2}) is the usual equation for mean
magnetic field evolution (see e.g. \cite{KR}) with $\alpha$-effect based on
particle effects rather on the averaging of turbulent pulsations. It is
well-known (see e.g. \cite{ZRS}) that Eq.~(\ref{Faradey2}) describes a
self-excitation of a magnetic field with the spatial scale $\Lambda\approx
\eta/\alpha $ and the growth rate $\alpha^2/4 \eta$.

Substituting $\alpha $ into $\Lambda =\eta /\alpha $ we arrive now to the
estimate 
\begin{equation}
\frac{\Lambda }{l_{H}}=1.6\times 10^{9}\left( \frac{T}{{\rm {MeV}}}\right)
^{-5}\left( \frac{\lambda _{{\rm {fluid}}}^{(\nu )}}{l_{\nu }(T)}\right)
(\mid \xi _{\nu _{e}}(T)\mid )^{-1}~,  \label{scale}
\end{equation}%
where the neutrino mean free path $l_{\nu }(T)=\Gamma _{W}^{-1}$ is given by
the weak rate $\Gamma _{W}=5.54\times 10^{-22}(T/{\rm MeV})^{5}~{\rm MeV}$, $%
l_{H}$$(T)=(2 {\cal {H})}$$^{-1}$ and ${\cal {H}}$=$4.46\times 10^{-22}(T/%
{\rm MeV})^{2}~{\rm MeV}$ is the Hubble parameter.

If the neutrino fluid inhomogeneity scale $\lambda^{(\nu)}_{{\rm fluid}}$ is
of the order $l_{\nu}(T_0)\sim 4~{\rm cm}\ll l_{H}(T_0)\sim 10^6~{\rm cm}$,
we have $\Lambda/l_{H}\geq 1$ at the beginning of the lepton era ($T=T_0\sim
10^2~{\rm MeV}$). The magnetic field time evolution is given by

\begin{equation}
B(t)=B_{\max }\exp \left( \int_{t_{\max }}^{t}\frac{\alpha ^{2}(t^{\prime })%
}{4\eta (t^{\prime })}dt^{\prime }\right) ~,  \label{dynamo}
\end{equation}%
where $B_{\max }$ is some seed value at the instant $T_{\max }\ll T_{EW}\sim
100$ {\rm GeV} (here we imbed the standard estimates of $\alpha ^{2}$
-dynamo into the context of expanding Universe and rely on the point-like
Fermi interaction used above).

For $\lambda _{{\rm fluid}}^{(\nu )}(T)\sim l_{\nu }(T)$ we can estimate the
index in the exponent (\ref{dynamo}) substituting in the integrand the
expansion time\newline
$t(T)=3.84\times 10^{21}(T/{\rm MeV})^{-2}{\rm MeV}^{-1}/\sqrt{g^{\ast }}$
with the effective number of degrees of freedom $g^{\ast }\sim 100$ at the
temperatures $T>1~{\rm GeV}$. Then from our estimates of $\alpha (T)$ and $%
\eta (T)$ with the change of the variable $(T/2\cdot 10^{4}{\rm MeV}%
)\rightarrow {\it x}$ one finds the fast growth of the mean field (\ref%
{dynamo}) in hot plasma, ${\it x}\leq 1$, 
\begin{equation}
B(x)=B_{\max }\exp \left( 25\int_{x}^{1}\left( \frac{\xi _{\nu _{e}}({\it {\
x^{\prime }})}}{0.07}\right) ^{2}{\it {x^{\prime }}^{10}d{x^{\prime }}}
\right)  \label{last}
\end{equation}%
given by the upper limit $x_{\max }=1,$ $T_{\max }=20~{\rm GeV}$. The
behaviour of $\xi _{\nu _{e}}(T)$ at high temperatures is unknown as well as
a value of the neutrino density asymmetry. We can state only that this value
changes due to neutrino oscillations somewhere below $T<10~{\rm MeV}$ not
overcoming the primordial nucleosynthesis limit $\mid \xi _{\nu _{e}}\mid
<0.07$ at the BBN time ($T\sim 0.1~{\rm MeV}$, $x=5\cdot 10^{-6}$) \cite%
{Dolgov1}. Nevertheless, even for $\mid \xi _{\nu _{e}}\mid \ll 0.07\ $
there remains an enhancement of a mean large-scale magnetic field $B_{\max
}\ll T_{\max }^{2}/\mid e\mid \ll T_{EW}^{2}/\mid e\mid $ by collective
neutrino-plasma interactions considered here, or this mechanism can be
efficient and important in cosmology. This is possible for a small neutrino
fluid inhomogeneity scale $\lambda _{{\rm fluid}}^{(\nu )}\ll l_{\nu }(T)$
entering the $\alpha $-parameter (\ref{helicity}) instead of the assumption $%
\lambda _{{\rm fluid}}^{(\nu )}\sim l_{\nu }(T)$ used in Eq. (\ref{last}).
Note that this free neutrino fluid parameter can vary in a wide region $%
T^{-1}\ll \lambda _{{\rm fluid}}^{(\nu )}\leq l_{\nu }(T).$

Let us remind that the inflation mechanism (with a charged scalar field
fluctuations at super-horizon scales) explains the origin of mean field at
cosmological scales. However, the value of this field is too small for
seeding the galactic magnetic fields. The amplification mechanism suggested
in our paper \cite{SS} can improve this very low estimate by a substantial
factor from Eq. (\ref{last}).

Thus, while in the temperature region $T_{EW}\gg T\gg T_0=10^2~{\rm MeV}$
there are many small random magnetic field domains, a weak mean magnetic
field turns out to be developed into the uniform {\it global} magnetic
field. The global magnetic field can be weak enough to preserve the observed
isotropy of cosmological model \cite{Zeld} while strong enough to be
interesting as a seed for galactic magnetic fields. This scenario was
intensively discussed by experts in galactic magnetism \cite{Kulsrud},
however until now no viable origin for the global magnetic field has been
suggested. We believe that the $\alpha^2$-dynamo based on the $\alpha$
-effect induced by particle physics \cite{SS} solves this fundamental
problem and opens a new and important option in galactic magnetism.

I thank Prof. D.D. Sokoloff for the illuminating discussions and the AHEP
group at IFIC guided by Prof. J.W.F. Valle for hospitality during part of
this work. This work was partially supported by the program of Presidium RAS
``Non-stationary phenomena in astronomy''.

\end{document}